# Gate-tunable *h/e*–period magnetoresistance oscillations in Bi$_2$O$_2$Se nanowires

Jianghua Ying[1,2], Guang Yang[1,2], Zhaozheng Lyu[1,2], Guangtong Liu[1], Zhongqing Ji[1], Jie Fan[1], Changli Yang[1], Xiunian Jing[1], Huaixin Yang[1,2], Li Lu[1,2,3,4,*] and Fanming Qu[1,3,4,*]

[1]Beijing National Laboratory for Condensed Matter Physics, Institute of Physics, Chinese Academy of Sciences, Beijing 100190, China
[2]School of Physical Sciences, University of Chinese Academy of Sciences, Beijing 100049, China
[3]CAS Center for Excellence in Topological Quantum Computation, University of Chinese Academy of Sciences, Beijing 100190, China
[4]Songshan Lake Materials Laboratory, Dongguan, Guangdong 523808, China
* Corresponding authors: lilu@iphy.ac.cn, fanmingqu@iphy.ac.cn.

**We report on the successful synthesis and low-temperature electron transport investigations of a new form of material - Bi$_2$O$_2$Se semiconducting nanowires. Gate-tunable 0- and π-*h/e* (*h* is the Planck constant and *e* the elementary charge) periodic resistance oscillations in longitudinal magnetic field were observed unexpectedly, demonstrating novel quasi-ballistic, phase-coherent surface states in Bi$_2$O$_2$Se nanowires. By reaching a very good agreement between the calculated density of states and the experimental data, we clarified the mechanism to be the one dimensional subbands formed along the circumference of the nanowire rather than the usually considered Aharonov-Bohm interference. A qualitative physical picture based on downward band bending associated with the complex band structure is proposed to describe the formation of the surface states.**

New-type low-dimensional materials have attracted tremendous interest due to their intriguing properties and application potentials in next-generation electronics, optoelectronics, etc. Graphene, phosphorene and transition-metal dichalcogenide are among the star materials[1-3]. However, drawbacks also exist in these materials, either lack of a bandgap, instable in air, or with only a moderate electron mobility. Recently, a new layered semiconductor Bi$_2$O$_2$Se drawn ever growing attention due to its superior electronic properties such as an ultrahigh electron mobility (~2.8 × 10$^5$ cm$^2$/V·s at 2 K), a tunable bandgap (~0.8 eV), and being stable in ambient environment[4-21]. Large scale Bi$_2$O$_2$Se thin films from monolayer to few-layer have been successfully synthesized through a chemical vapor deposition method[4,17]. The existence of strong spin-orbit interaction[21], and high-performance field-effect transistors and optoelectronics[4,11,12,15,19,22] have also been experimentally demonstrated. And the ferroelectricity/ferroelasticity is further expected[8].

Given the fascinating features in the 2D form of Bi$_2$O$_2$Se, it is natural to expect that the 1D nanowire form of Bi$_2$O$_2$Se, with large surface-to-volume ratio, would retain some of the unique properties of its 2D counterpart. However, investigation on Bi$_2$O$_2$Se nanowires is still a mysterious veil at present. In this work, we report on the synthesis of high-quality single-crystalline Bi$_2$O$_2$Se nanowires by means of gold-catalyzed vapor-liquid-solid (VLS) mechanism. Low-temperature electron transport measurements reveal $h/e$–period resistance oscillations as a function of longitudinal magnetic field ($h$ is the Planck constant and $e$ the elementary charge) and a gate-tunable π-phase shift. These unexpected observations demonstrate quasi-ballistic, phase coherent electron transport of surface states in $Bi_2O_2Se$ nanowires. A very good agreement between the calculated density of states and the experimental results – the double modulation of the resistance oscillations by gate voltage and magnetic field, clarifies the mechanism to be the 1D subbands formed along the circumference of the nanowire rather than the usually considered Aharonov-Bohm interference. A qualitative picture associated with the complex band structure is further proposed to interpret the formation of surface electron accumulation based on downward band bending.

$Bi_2O_2Se$ nanowires were synthesized via gold-catalyzed VLS mechanism on Si wafers in a horizontal tube furnace with $Bi_2Se_3$ powder as the source material. A detailed phase equilibrium process can be found in Supplemental Material[23]. After growth, the wafer appeared to be gray as an evidence of nanowire production. Figure 1a shows the scanning electron microscopy (SEM) image of the as-synthesized $Bi_2O_2Se$ nanowires, which have a length of 5 - 10 μm and a diameter of 80 - 250 nm.

To characterize the basic crystal structure, we have carried out an extensive transmission electron microscopy (TEM) study on the as-synthesized samples. Figure 1b shows a bright-field TEM image and the corresponding diffraction pattern taken along the [010] zone axis direction of one typical nanowire. All the main diffraction spots in these patterns can be well indexed using the expected tetragonal unit cell with lattice parameters of about $a$= 0.389 nm, $c$=1.22 nm (space group of I4/mmm)[4], which agree well with the X-ray powder diffraction result as shown in Supplemental Material[23]. A better and very clear view of the atomic structure for the various positions has been obtained by high-resolution TEM (HRTEM) observations. Figures 1c and 1d show the HRTEM images taken from the square-enclosed A and B regions in Fig. 1b, respectively. It can be seen from Fig. 1b that the $Bi_2O_2Se$ nanowire attaches with one well-defined facet of gold nano-particle and grows along the [001] direction. From Fig. 1b we can conclude that there is an angle between the [001] atomic layer and the growth direction (the long axis direction) of the nanowire.

The HRTEM images in Fig. 1b also indicate the presence of zigzag structure along the edge of the as-prepared nanowires. Figure 1d illustrates the HRTEM image taken from region B, demonstrating that there is no obvious planar defect in such areas. The layered structural morphologies are very likely caused by a layer-by-layer growth mode, in which there could be a relative shift in the *ab* plane during the crystal growth due to the specific synthesis condition. These spectroscopic characterizations demonstrate that the as-synthesized $Bi_2O_2Se$ nanowires are of high quality.

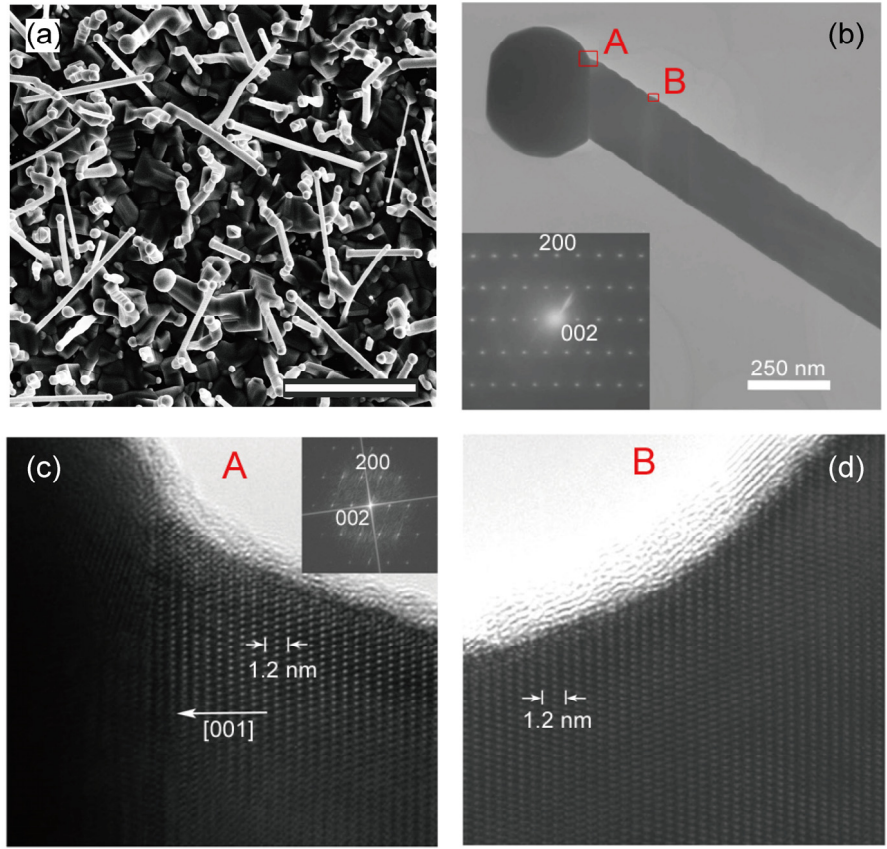

Figure 1. (a) SEM image of the as-synthesized $Bi_2O_2Se$ nanowires on Si wafer. Scale bar: 5 μm. The ball on the top of the nanowire is the Au catalyst. (b) A bright-field TEM image and the corresponding diffraction pattern taken along the [010] zone axis direction of one typical nanowire. (c) and (d) The HRTEM images taken from the square-enclosed A and B regions in (b), respectively. The inset in (c) presents the FFT spectrum.

For electron transport measurements, $Bi_2O_2Se$ nanowires were mechanically transferred onto a highly doped Si substrate with a 300 nm thick $SiO_2$ layer used for applying a back-gate voltage. Ti/Au (5 nm/150 nm) contacts were deposited using standard lithography technique. Two-terminal measurements were carried out with a temperature down to 1.8 K and a magnetic field up to 7 T.

Figure 2a shows a sketch of the measurement setup. The large surface-to-volume ratio of the nanowire provides excellent geometries for probing the transport properties of surface states, if present, as indicated by the red arrows in Fig. 2a. Figure 2b presents an AFM image of device #1 with two bright Ti/Au contacts. Figure 2c displays the two-terminal resistance versus magnetic field *B* applied along the axis of the nanowire and back-gate voltage $V_g$ at 1.8 K for device #1. In order to identify the oscillating features more clearly, we smoothed the data of Fig. 2c and extracted the second order derivatives, i.e., the phase diagram, as shown in Fig. 2d.

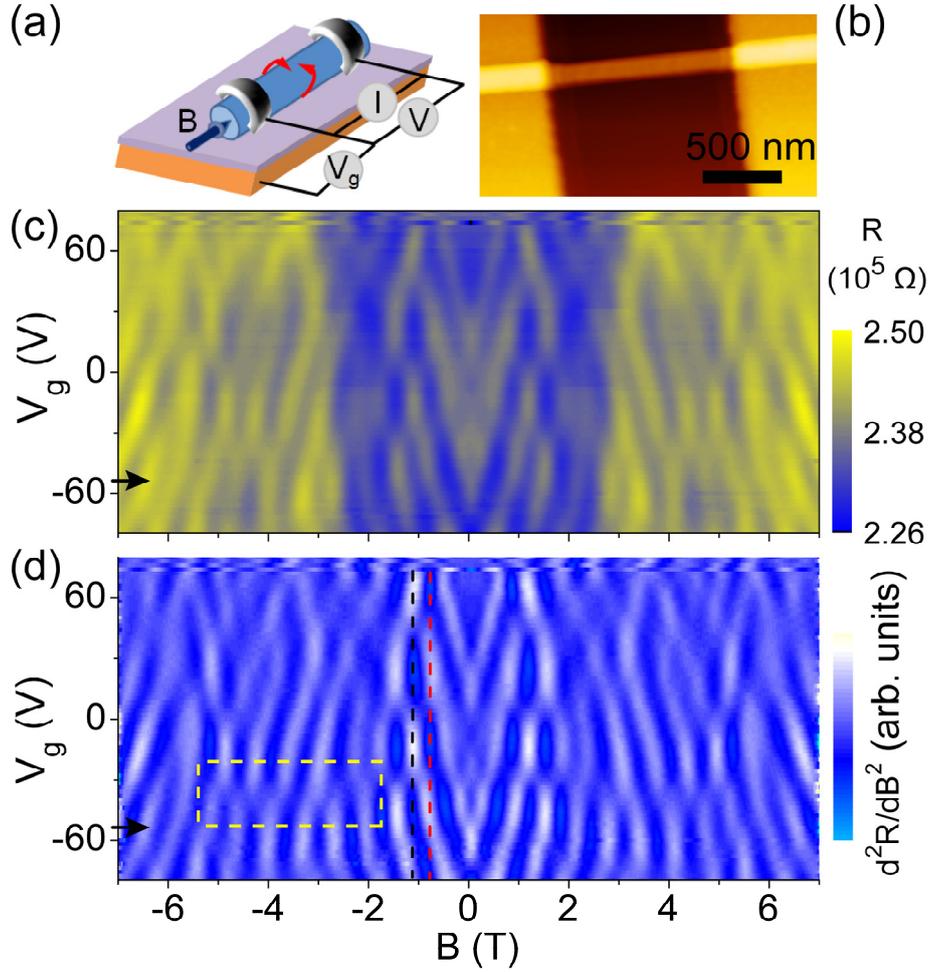

Figure 2. (a) Sketch of two-terminal measurement configuration. A longitudinal magnetic field $B$ and a back-gate voltage $V_g$ are applied. (b) Atomic force microscopy (AFM) image of device #1 with two Ti/Au contacts. (c) Color map of the two-terminal resistance $R$ versus $B$ and $V_g$ at 1.8 K. (d) Second order derivatives of the smoothed data from (c), clearly illustrating the oscillations. The yellow rectangle region highlights the π-phase shifts. The black and red dashed lines mark the two line cuts in Fig. 3e. The arrows in (c) and (d) indicate the positions of the line cuts in Figs. 3a and 3c, respectively.

One obvious characteristic of Figs. 2c and 2d is that the resistance oscillates as a function of $B$. The periodic oscillations in magnetic field along the nanowire axis (see representative curves in Figs. 3a and 3c) indicate a well-defined surface conducting channel to pick up flux[24]. Such periodic oscillations with a period of $h/e$ or $h/2e$ have been observed in many 1D systems with a surface conducting layer[24], such as metallic cylinders[25], carbon nanotubes[26], core/shell nanowires[27,28], semiconducting nanowires[29], topological insulator nanoribbons[30,31], Dirac semimetal nanowires[32], etc. Note that only a few small band gap semiconductors carry intrinsic electron-accumulated surface states, such as InN, InAs and ZnO[33-35]. In addition, persistent currents in normal-metal rings parallels with equivalent results[36-38].

In a mesoscopic ring, the mechanism of periodic resistance oscillations is relatively straightforward. The *h*/2*e*–period oscillations, named as Altshuler-Aronov-Spivak oscillations[39], result from the interference of partial waves between a pair of time-reversed paths, sharing the same mechanism as weak localization. On the other hand, the *h*/*e*–period oscillations are associated with the interference of partial waves through two different possible paths from one terminal to the other one, called Aharonov-Bohm (AB) oscillations[40].

For cylindrical shell conductors, *h*/2*e* oscillations are usually observed at low magnetic fields, and die out and disappear at high magnetic fields due to the requirement of time-reversal symmetry[41]. On the contrary, the *h*/*e* oscillations can survive at both high and low magnetic fields, without the need of time-reversal symmetry. The periodic oscillations in Figs. 2c and 2d survive at high magnetic fields (±7 T), and thus they are *h*/*e* oscillations. However, the *h*/*e* oscillations require ballistic transport[28,29]. In such case, although it is still called AB oscillation[40], the underlying mechanism transits to the modulation of density of states by magnetic field rather than AB interference. The ballistic transport of electrons along the circumference forms 1D subbands with different angular momentum which is regulated by the penetrated magnetic field[28,29,42,43], as explained later. The competition between the *h*/*e* and *h*/2*e* components depends on the strength of disorder and the size of the cylinder[29,32,44].

To analyze the oscillations as a function of longitudinal magnetic field, we take typical line cuts at $V_g$=-54 V from Figs. 2c and 2d (marked by the black arrows), as shown in Figs. 3a and 3c, respectively. The periodic oscillations can be clearly recognized. Figure 3b shows the fast Fourier transform (FFT) spectrum of the *R* vs. *B* curve in Fig. 3a. A prominent peak can be found at a frequency of *f*=1.46 T$^{-1}$ as depicted by the triangle. The low-frequency peaks correspond to the large-scale background variation of Fig. 3a, presumably due to universal conductance fluctuations (UCFs) of the bulk states. These false peaks can be readily suppressed by taking the FFT spectrum of the second order derivative d$^2$*R*/d*B*$^2$, a method commonly applied to separate the oscillatory part from background, as shown in Fig. 3d. The *h*/*e* periodic oscillations are thus associated with an area of *S*=*f*×*h*/*e*=6050 nm$^2$, where *S* is the cross-section area of the nanowire. Taking the nearly elliptical cross-section dimensions, a width of 140 nm and a height of 95 nm, we need to assume a depth *d* of the surface states[30] to match the required area. *d*~6 nm was deduced from π×(65-*d*)×(42.5-*d*)=*S* after subtracting the surface undulation of 5 nm due to zigzag structures. Note that the dimensions might be overestimated from SEM or AFM images. Nevertheless, the *h*/2*e* oscillations at a double frequency 2*f* were invisible. Combining the dominant *h*/*e* oscillations and the superposition of UCFs, we conclude that our Bi$_2$O$_2$Se nanowires are in the quasi-ballistic regime[27,29,45].

Another noticeable characteristic of Figs. 2c and 2d is that the resistance changes between a peak and a valley alternatively along gate-voltage (even without magnetic field). Figure 3e depicts two curves of d$^2$R/dB$^2$ vs. $V_g$ taken out from Fig. 2d at B=-1.13 T and -0.73 T, as marked by the black and red dashed lines, respectively. The oscillating behavior arises from the variation of the density of states as tuning the Fermi level by $V_g$ (as explained later). The opposite phase of these two oscillating curves demonstrates a relative π-phase shift. Under the double modulation by $V_g$ and B, the π-phase shift along $V_g$ also represents tuning between the 0- and π-h/e oscillations along B, i.e., starting from low and high resistance at zero magnetic field, respectively. Such π-phase shift along B is more obvious in the dashed rectangle in Fig. 2d.

In the following we explain the above observations consistently within the theoretical framework of ballistic electron transport through cylindrical surface states penetrated by a magnetic flux[28,29,32,46]. The motion of surface electrons along the cylindrical shell can be regarded as a combination of a circular component and a free component along the nanowire axis z. The Hamiltonian can be written as

$$\hat{H} = \frac{1}{2m^* r_0^2} (L_z - \frac{1}{2} eB_z r_0^2)^2 - \frac{\hbar^2}{2m^*} \frac{\partial^2}{\partial z^2}$$

where $L_z$ denotes the angular momentum operator and $r_0$ the radius of the nanowire. With π$r_0^2 B_z$ =Φ as the magnetic flux through the circular loop and $\Phi_0 = h/e$, we find the energy eigenvalues as

$$E_l = \frac{\hbar^2 k_z^2}{2m^*} + \frac{\hbar^2}{2m^* r_0^2} (l - \frac{\Phi}{\Phi_0})^2$$

The first term describes the kinetic energy for the motion along z. The second term, where angular momentum quantum number $l = 0, \pm 1, \pm 2$ ...., represents a series of parabolas as a function of Φ shifted to each other by $\Phi_0$. The energetic spectrum for $k_z$=0 ($E_l'$) with each parabola belonging to a specific angular momentum state is illustrated in Fig. 3f by the solid gray lines. However, in contrast to the experimental observations shown in Figs. 2c and 2d, h/e-periodic oscillations along B are only expected at certain discrete energies, as marked by the black bars in Fig. 3f. And irregular oscillations dominate for most of the energies as represented by the horizontal green dashed line in Fig. 3f. We argue that level broadening of the subbands presumably arising from thermal broadening, inelastic scattering and/or finite thickness of the surface channel (uncertainty of flux) should be taken into account.

From the red line of Fig. 4e, we take one full oscillation from $V_g$=-64 V to 12 V to estimate the Fermi level position. Under the assumption of finite bulk contributions, this gate-voltage range corresponds to a Fermi level shift of 13.2 meV (subband separation) [23]. Using $E_l'(\Phi = 0) = \frac{\hbar^2}{2m^* r_0^2} l^2$, we extract the Fermi level at $V_g$=-64 V to be 299.4 meV ($l = \pm 46$), at $V_g$=12 V to be 312.6 meV ($l = \pm 47$). The colored background of Fig. 3f depicts the calculated density of states (DOS) by approximating level broadening as a Lorentz shape distribution. Clearly, the double modulation by $B$ and $V_g$ gives rise to the $\Phi_0$-period DOS (resistance) oscillations along $B$ as well as the π-phase shift. When $V_g$ tunes the Fermi level, the DOS (resistance) also oscillates at a given $B$. The opposite phase between the black and red dashed lines can be apparently recognized, as shown by the two experimental curves in Fig. 3e accordingly. The π-phase shift, i.e., the 0- and π-$h/e$ oscillations, as marked by the black (DOS peak at $\Phi=0$) and red (DOS valley at $\Phi=0$) arrows in Fig. 3f, respectively, can be immediately identified. Note that the DOS map in Fig. 3f represents a qualitative interpretation of the data. Details of the extraction of the DOS color map can be found in the Supplemental Material[23].

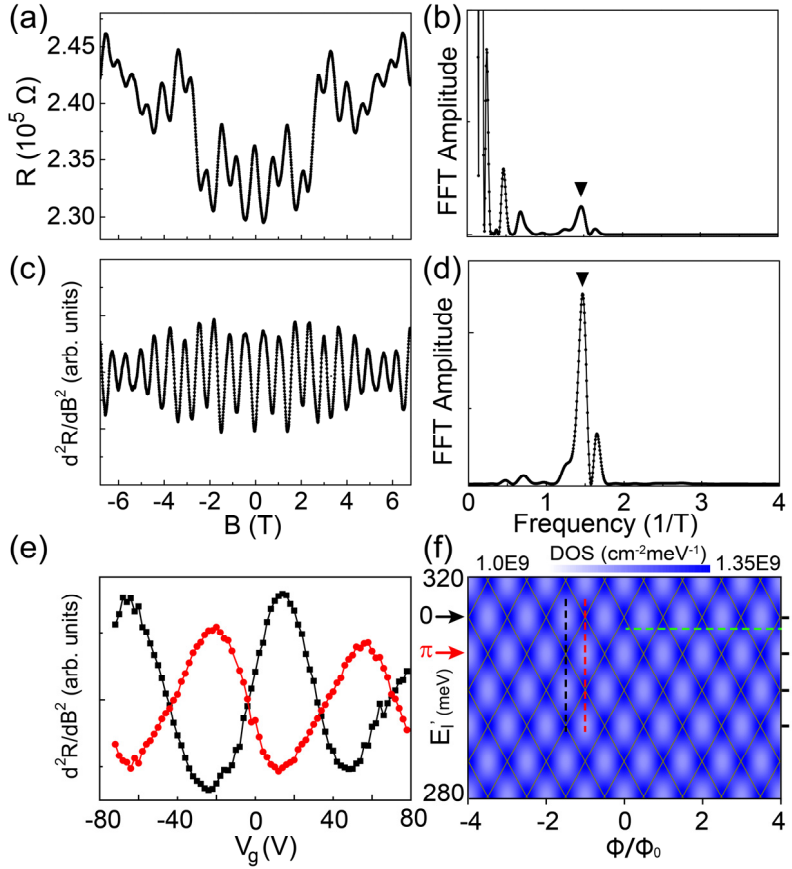

Figure 3. (a) $R$ vs. $B$ curve at $V_g$=-54 V taken out from Fig. 2c. (b) FFT of $R$ vs. $B$ curve in (a). (c) Second order derivative d$^2R$/d$B^2$ of the smoothed $R$ vs. $B$ curve in (a) and its FFT pattern is shown in (d). The triangles in (b) and (d) denote the main peak at 1.46 T$^{-1}$. (e) Two line cuts of d$^2R$/d$B^2$ vs. $V_g$ taken out from Fig. 2d as marked by the black ($B$=-1.13 T) and red ($B$=-0.73 T) dashed lines, respectively. (f) Subbands formed at the circumference of the nanowire as a function of longitudinal magnetic flux $\Phi$, without level broadening (solid gray lines) and with level broadening (DOS represented by bluish background colors). The green dashed line indicates $\Phi_0$-period (irregular) oscillations with (without) level broadening, while the bars show $\Phi_0$-period oscillations for both cases. A π-phase shift exists both in the $V_g$ dependence of oscillations (i.e., along the dashed black and red line cuts), and in the $\Phi$ dependence of oscillations (i.e., line cuts along the black and red arrows).

Furthermore, we have performed systematic measurements of gate-dependent $h/e$ oscillations on device #2. Consistent oscillations as a function of both $B$ and $V_g$ were observed, as shown in Figs. 4a and 4b, including the alternation between resistance peaks and valleys at $B$=0 and the π-phase shift (marked by the dashed line in Fig. 4b).

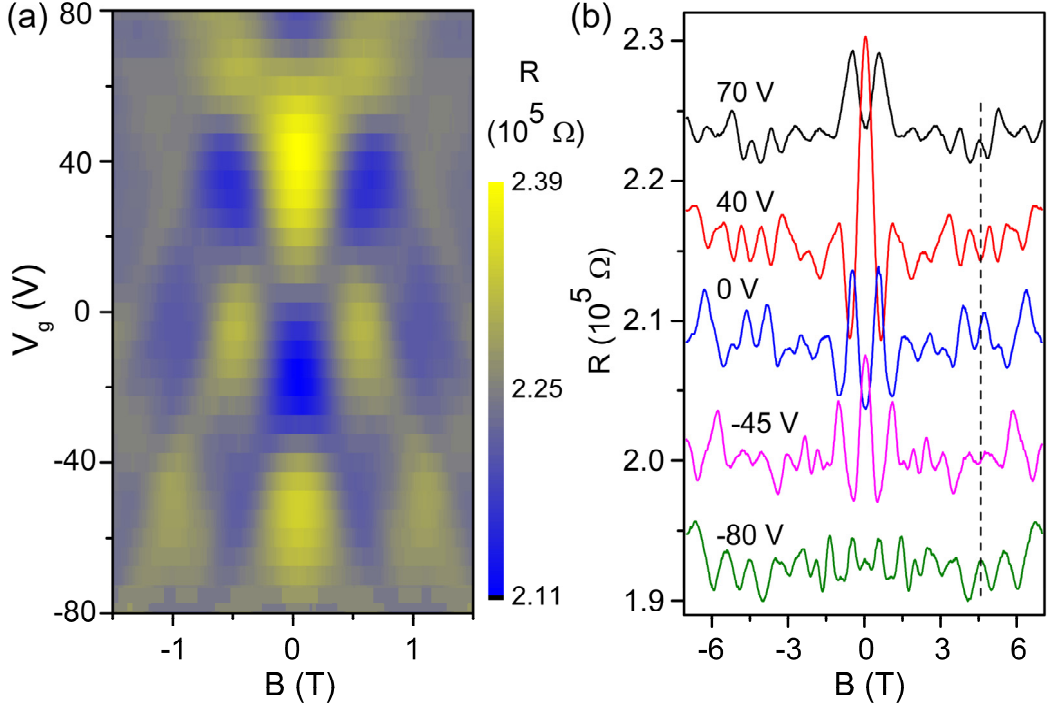

Figure 4. (a) Resistance $R$ of device #2 as a function of both $B$ and $V_g$ at 1.8 K. (b) $R$ vs. $B$ curves at different $V_g$ with an offset for clarity. The black dashed line guides the π-phase shift between different curves.

Next, we propose a qualitative physical picture to describe the formation of surface conducting channels on $Bi_2O_2Se$. For semiconductors, the surface Fermi level $E_F$ tends to sit at a universal value of ~4.9 eV[47] or ~4.5 eV (with hydrogen adsorbed)[33] below the vacuum level to reach charge neutrality. This charge neutrality level, called branch point energy ($E_B$), represents the separation of donor-type (below) and acceptor-type (above) states at the surface/interface[48]. It also corresponds to the average mid-gap energy weighting the entire Brillouin zone in the complex band structures[49]. For a few small band gap semiconductors with an extremely low conduction band minimum ($E_C$) and a small effective mass, the density of states around Γ point is negligible compared to other band edges at the boundaries of the Brillouin zone. Therefore, average mid-gap energy $E_B$ could be higher than $E_C$, such as in InN, InAs and ZnO[33-35]. The direct result is the existence of donor-type states at the surface, and thus the downward band bending and accumulation of electrons near the surface.

Regarding $Bi_2O_2Se$, the band structures indeed show a low $E_C$ at Γ point, ~0.8 eV above the valence band maximum $E_V$, with a relative small effective mass of $0.14m_0$ ($m_0$ is the free electron mass)[4]. Consequently, it is possible that $E_B$ is higher than $E_C$. Applying similar approximations as for small gap III-V semiconductors[49], the mid-gap energy at *X* point, the zone face center, indeed crosses the conduction band at Γ point[4,7,8,10,50]. Due to the variation between the band structures of $Bi_2O_2Se$ obtained by different methods[4,7,8,10,15,50-52], however, we seek to propose a possible qualitative picture in the current work. Assuming that $E_B$ is higher than $E_C$ and a position of $E_F$ in the gap, the band alignment near the surface is illustrated in Fig. 5. The unoccupied donor-type states at the surface are positively charged, balanced by the downward band bending induced space charge. Near the surface, $E_F$ crosses $E_C$, and thus electrons accumulate to form surface conducting channels. Note that this mechanism originates from the intrinsic donor-type states at the surface. Further detailed investigations are required to shed more light on the surface states of $Bi_2O_2Se$.

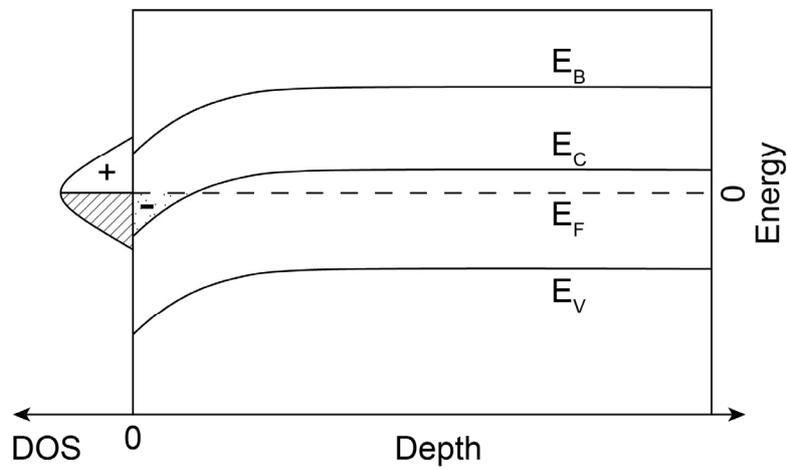

Figure 5. The branch point energy $E_B$ is assumed to be higher than the conduction band minimum $E_C$. The left side of the figure shows the density of states of the donor-type states at the surface, which are positively charged when unoccupied. The band bends downward to reach charge neutrality, resulting in the crossing of Fermi level $E_F$ with conduction band and thus electron accumulation near the surface.

In summary, we synthesized high quality $Bi_2O_2Se$ nanowires by means of a gold-catalyzed VLS method. The single-crystalline nature of $Bi_2O_2Se$ nanowire as a typical bismuth-based oxychalcogenide material was confirmed by X-ray diffraction and transmission electron microscopy. The gate-tunable *h/e*–period oscillations were observed and attributed to the quasi-ballistic transport through the surface states of $Bi_2O_2Se$ nanowires. These results clearly demonstrate a new type of semiconducting nanowires with quasi-1D electron subbands of different angular momentum at the surface, which provide an ideal platform for the design of future quantum electronics. A possible intrinsic mechanism is proposed for the formation of electron accumulation, which, however, deserves further detailed theoretical and experimental inspections.

**Acknowledgments**：

We would like to thank Yugui Yao, Xuefeng Wang, Fan Yang, Jun Chen and Yongqing Li for fruitful discussions. This work was supported by the National Basic Research Program of China from the MOST grants 2017YFA0304700, 2016YFA0300601, and 2015CB921402, by the NSF China grants 11527806, 91221203, 11174357, 91421303, 11774405, by Open Research Fund from State Key Laboratory of High Performance Computing of China, by Beijing Academy of Quantum Information Sciences, Grant No. Y18G08, and by the Strategic Priority Research Program B of Chinese Academy of Sciences, Grant No. XDB28000000, XDB07010100.

# Supplemental Material for

# Gate-tunable *h/e*–period magnetoresistance oscillations in Bi$_2$O$_2$Se nanowires

Jianghua Ying[1,2], Guang Yang[1,2], Zhaozheng Lyu[1,2], Guangtong Liu[1], Zhongqing Ji[1], Jie Fan[1], Changli Yang[1], Xiunian Jing[1], Huaixin Yang[1,2], Li Lu[1,2,3,4,*] and Fanming Qu[1,3,4,*]

[1]*Beijing National Laboratory for Condensed Matter Physics, Institute of Physics, Chinese Academy of Sciences, Beijing 100190, China*
[2]*School of Physical Sciences, University of Chinese Academy of Sciences, Beijing 100049, China*
[3]*CAS Center for Excellence in Topological Quantum Computation, University of Chinese Academy of Sciences, Beijing 100190, China*
[4]*Songshan Lake Materials Laboratory, Dongguan, Guangdong 523808, China*
[*] Corresponding authors: lilu@iphy.ac.cn, fanmingqu@iphy.ac.cn.

**Growth of Bi$_2$O$_2$Se nanowires.** Bi$_2$O$_2$Se nanowires were synthesized in a horizontal tube furnace via gold-catalyzed vapor-liquid-solid mechanism (Fig. S1). Firstly, Si wafers were coated with AuCl$_3$·HCl·4H$_2$O solution. Secondly, the wafers were annealed at 700 ℃ for an hour with 200 sccm H$_2$ in the tube furnace to decompound AuCl$_3$·HCl·4H$_2$O and obtain Au nanoparticles with sizes ranging from 50 to 300 nm. The wafers were then ready for nanowire growth. About 2 g Bi$_2$Se$_3$ powder (99.999%) was put at the center of the furnace, and the Si wafer was put at the downstream with a distance of 12 - 14 cm. During the growth, a 400 - 500 sccm Ar carrier gas and a 6000 - 8000 Pa pressure were maintained. Note that due to the tiny requirement of O$_2$ for growing the near zero volume Bi$_2$O$_2$Se nanowires, the residual and leakage O$_2$ is enough without the necessity of extra O$_2$ supply. (When ~40 sccm O$_2$ is supplied, no production is obtained.) A detailed phase equilibrium process can be found in Fig. S2. The temperature at the center of the furnace was typically 580 ℃ and the temperature of the wafer was 500 - 530 ℃. The growth time was 5 - 10 h.

**Electron transport measurement.** For electron transport measurements, Bi$_2$O$_2$Se nanowires were mechanically transferred from Si wafer onto a highly doped Si substrate with a 300 nm thick SiO$_2$ layer used for applying a back gate voltage. Electrode patterns were fabricated by electron-beam lithography, and Ti/Au (5 nm/150 nm) contacts were deposited by electron-beam evaporation. Two-terminal measurements were carried out using standard lock-in technique with a temperature down to 1.8 K and a magnetic field up to 7 T.

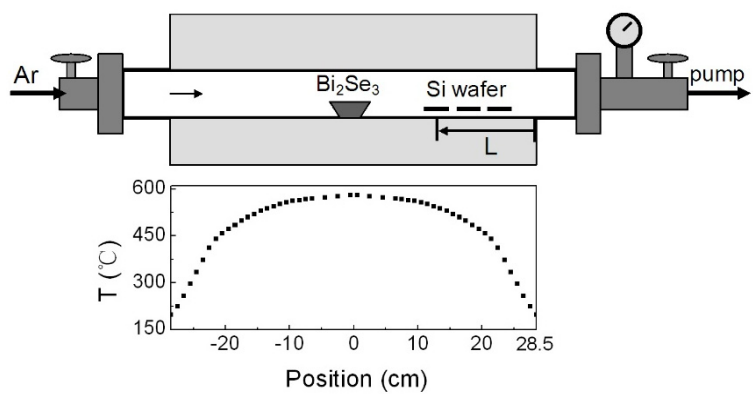

**Figure S1.** The edge of the furnace is at the position of ±28.5 cm. During growth, the temperature at the center of the furnace is kept at 580 °C. Bi$_2$Se$_3$ powder is put at the center of the furnace and Si wafers with Au catalyst are put at the downstream with a distance L to the edge of the furnace.

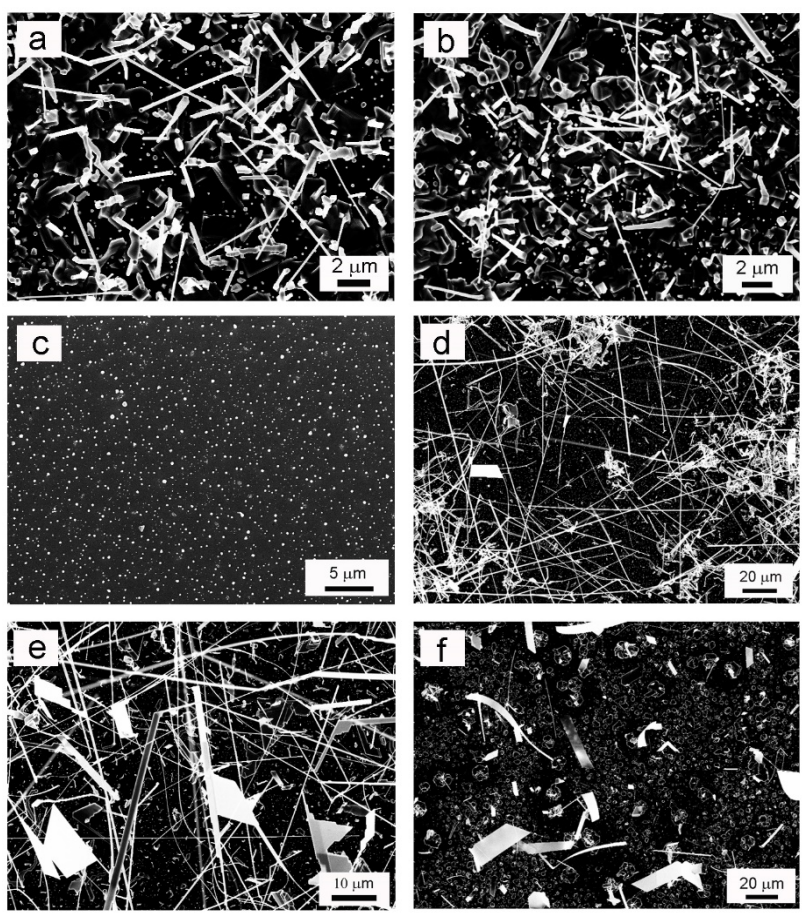

**Figure S2.** (a) *L*=13 cm, *T*=520 °C (see the meaning of *L* in Fig. S1). (b) *L*=12 cm, *T*=510 °C. (c) *L*=10 cm, *T*=485 °C. (d) *L*=8 cm, *T*=457 °C. (e) and (f) *L*=6 cm, *T*=412 °C. Only the Si wafer of (f) had no Au catalyst. The productions: (a) and (b) Bi$_2$O$_2$Se nanowires, (c) no deposition but Au catalyst, (d) mainly Bi$_2$Se$_3$ nanoribbons, (e) Bi$_2$Se$_3$ nanoribbons and nanoplates, (f) mainly Bi$_2$Se$_3$ nanoplates. The results show that the productions depend on the temperature of the wafer and Bi$_2$Se$_3$ can be obtained without Au catalyst.

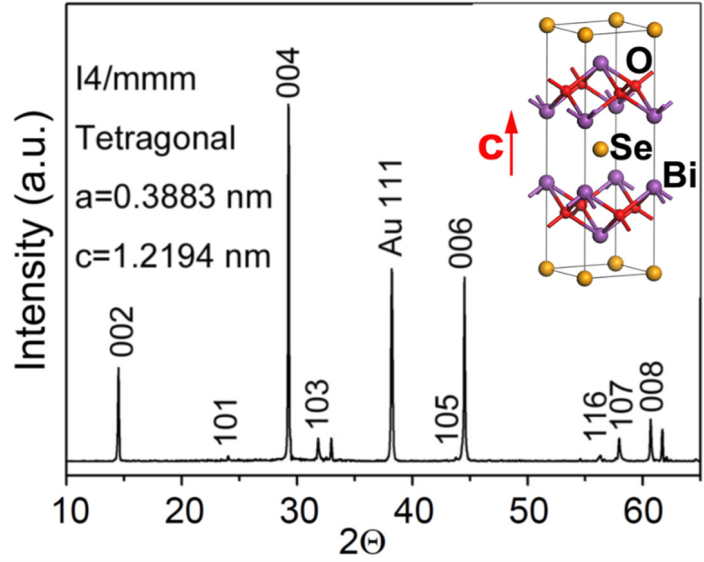

**Figure S3.** The X-ray powder diffraction pattern taken from a typical Bi$_2$O$_2$Se sample grown on Si wafer. Most of the diffraction peaks can be indexed using the tetragonal unit cell with lattice parameters of *a*=0.3883 nm, *c*=1.2194 nm (space group of I4/mmm). The inset depicts a schematic structural model of the Bi$_2$O$_2$Se crystal, clearly illustrating the alternating sequence of the Bi-O layers and Se layers stacking along the *c*-axis direction.

**Calculation of the density of states (DOS):**

The solid lines in Fig. 3f in the main text represent the subbands with different angular momentum for $K_Z$=0 following the formula

$$E'_l = \frac{\hbar^2}{2m^* r_0^2}(l-\frac{\Phi}{\Phi_0})^2.$$

However, due to thermal widening, inelastic scattering and the finite thickness of the surface channels, level broadening instead of DOS singularities should be considered.

We show the detailed calculations below.

**1)** The dependence on gate-voltage can help to attain the information about the change of the carrier density, thereby further determine the location of the Fermi energy.

We assume that the back-gate voltage affects the whole nanowire uniformly (as an estimate). For the 2D surface of the nanowire,

$$DOS(E) = \frac{dN}{dE} = \frac{m^*S}{\pi\hbar^2}$$

$$N = \int_0^{E_f} DOS(E)dE = \frac{m^*S}{\pi\hbar^2}E_f$$

$$n = N/S = \frac{m^*}{\pi\hbar^2}E_f$$

From the red line of Fig. 4e, we take one full oscillation from $V_g$=-64 V to 12 V to estimate the Fermi level position. As $Bi_2O_2Se$ is usually a n-type semiconductor, we assume a Fermi level of ~200 meV above the conduction band for the bulk [1]. As a result, this gate-voltage range corresponds to a Fermi level shift of 13.2 meV (subband separation).

Using $E_l'(\Phi = 0) = \frac{\hbar^2}{2m^*r_0^2}l^2$, we extract the Fermi level at $V_g$=-64 V to be 299.4 meV ($l = \pm 46$) above the bottom of the surface subband, at $V_g$=12 V to be 312.6 meV ($l = \pm 47$). At $V_g$=-64 V, we divide DOS(E) into the 92 subbands (a factor of 2 from orbital degeneracy) equally to calculate the DOS for each subband.

But in general, all of these calculations can be regarded as a rough estimate.

**2)** We consider the level broadening along the flux axis and the energy axis separately first.

We calculate the DOS using Lorentzian and Gaussian broadening and compare the amplitude of DOS oscillations at $E_l'(\Phi = 0) = 299.4$ meV ($l = \pm 46$) with the measured resistance oscillations around $V_g$=-64 V to determine the best fit.

Lorentzian: $A^2/[A^2 + (\Delta\Phi/\Phi_0)^2]$

Gaussian: $\frac{1}{\sqrt{2\pi}\sigma}\exp(-\frac{1}{2\sigma^2}(\frac{\Delta\Phi}{\Phi_0})^2)$

It turned out that they can both generate consistent DOS results to explain our experimental data. The general difference is just the line shape of the broadening. The results are: (1) For Lorentzian broadening along flux axis, $A$=0.8 fits the maximum resistance oscillation amplitude of ~3% the best. (2) For Lorentzian broadening along energy axis, $A$=10.2 fits the best. (3) For Gaussian broadening along flux axis, σ=0.5 fits the best. (4) For Gaussian broadening along energy axis, σ=6.5 fits the best.

We can see that it requires very huge broadening along energy to fit the data, while the broadening along flux is moderate. Both thermal broadening and scattering introduce smearing in energy, and finite thickness of the surface states induces broadening in flux. However, the broadening along energy seems not capable to fit the data. If we take Lorentzian for an example, $A$=10.2 means a full width at half maximum of 20.4 meV. A temperature of $T$=2 K corresponds to $K_BT$=0.17 meV. Although it is difficult to determine the broadening due to scattering, the quasi-ballistic nature suggests a small scattering rate. So that a 20.4 meV broadening seems unreasonable.

Below are the calculated DOS maps.

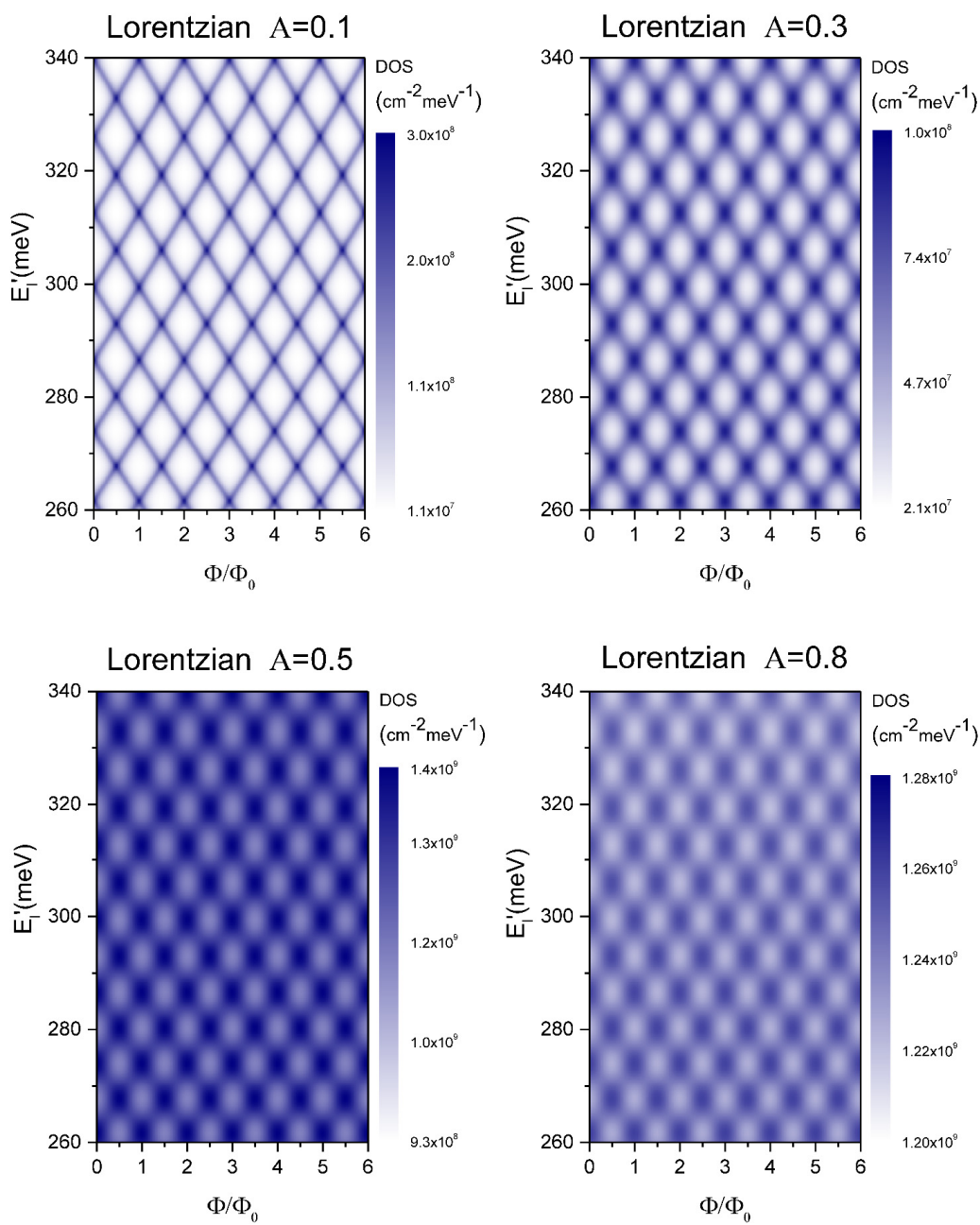

**Figure S4.** The figures above show the DOS maps for Lorentzian broadening along the flux axis with different broadening strength (*A*). The DOS map with *A*=0.8 fits the data best.

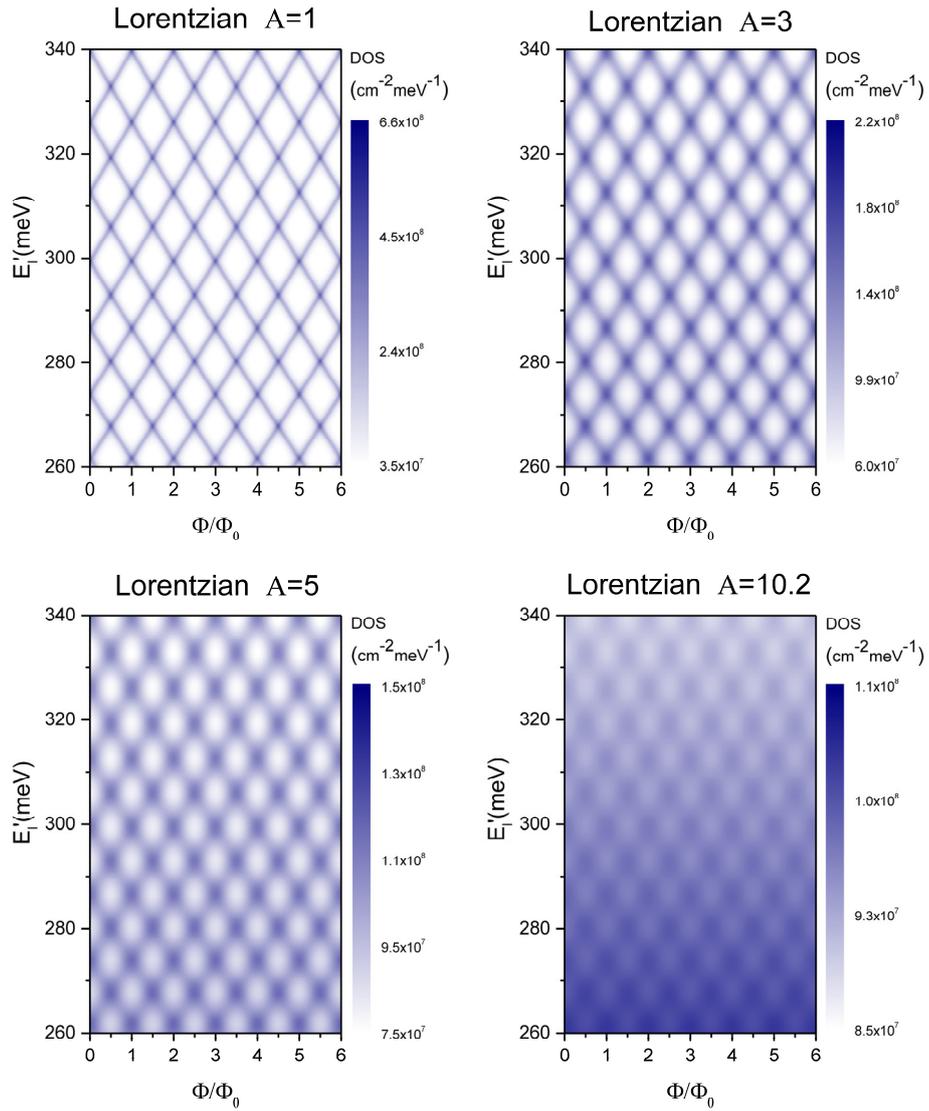

**Figure S5.** The figures above show the DOS maps for Lorentzian broadening along the energy axis with different broadening strength (*A*). The DOS map with *A*=10.2 fits the data best.

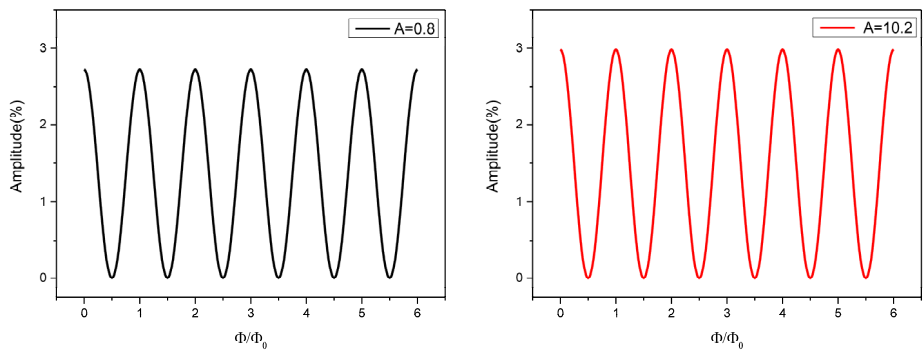

**Figure S6.** The figures above show a comparison between Lorentzian broadening along the flux axis (left, *A*=0.8) and along the energy axis (right, *A*=10.2) at $E_l'(\Phi=0) = 299.4$ meV ($l = \pm 46$). The amplitude of the DOS oscillations in both figures is close to 3%, the oscillation amplitude of the resistance around $V_g$=-64 V.

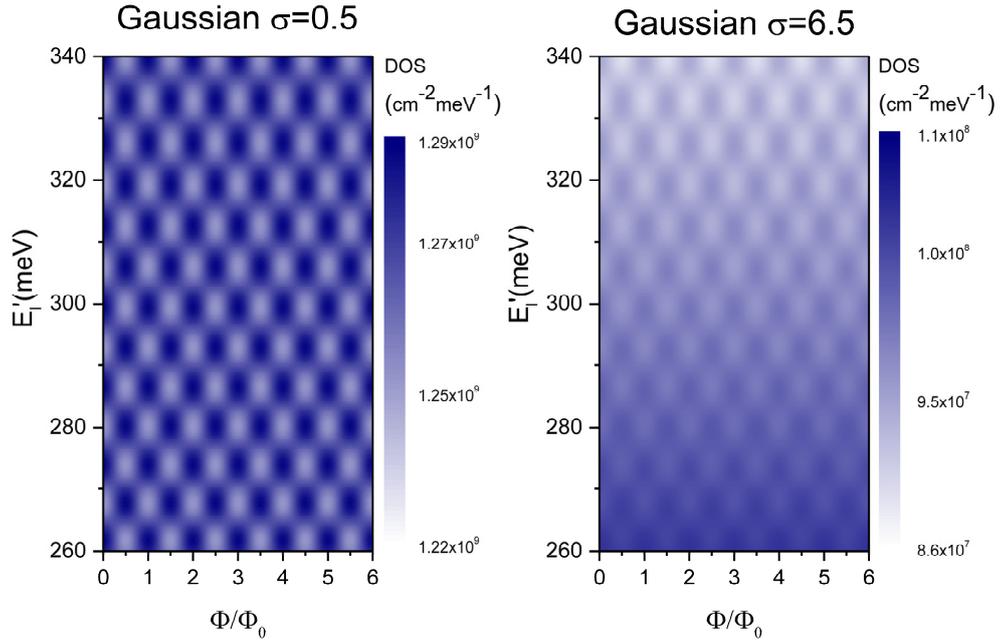

**Figure S7.** The figures above show a comparison between Gaussian broadening along the flux axis (left, σ=0.5) and along the energy axis (right, σ=6.5). At $E_l'(\Phi=0)=299.4$ meV ($l=\pm 46$), the amplitude of the DOS oscillations is close to 3%, the oscillation amplitude of the resistance around $V_g$=-64 V.

**3)** Broadening along flux and energy should coexist. We use a 2D Lorentzian $\dfrac{A_1^2 A_2^2}{(A_1^2+(\Delta\Phi/\Phi_0)^2)(A_2^2+(\Delta E)^2)}$ to include both contributions to the broadening of DOS. A series of ($A_1$, $A_2$) (broadening along flux and energy, respectively) fit the data well. For instance, ($A_1$, $A_2$)=(1.2, 14) or (1, 17). However, we cannot determine the ratio of the broadening along the two axes. In general, the energy broadening needs to be much larger than the flux broadening to fit the data.

Below are the calculated DOS maps.

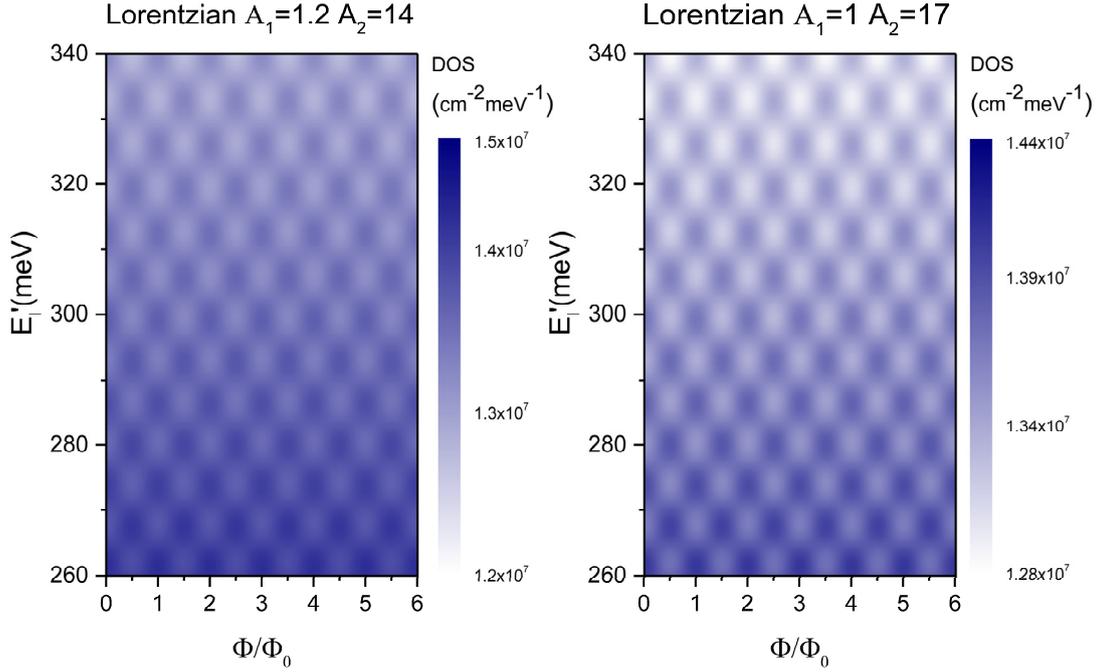

**Figure S8.** The figures above show a comparison between 2D Lorentzian broadening with different parameters ($A_1$, $A_2$). At $E'_l(\Phi = 0) = 299.4$ meV ($l = \pm 46$), the amplitude of the DOS oscillations is close to 3%, the oscillation amplitude of the resistance around $V_g$=-64 V.

**4)** After calculating the DOS using 1D and 2D functions, it turned out difficult to determine quantitative contributions from flux broadening and energy broadening. Given the fact that energy broadening needs to be (unlikely) huge, we decided to describe the broadening in a qualitative way, only considering the contribution of flux broadening. From the information above, we know that $A$=0.8 fits the best. However, energy broadening should also contribute, so that we chose a smaller $A$=0.5 to show the explanation using level broadening qualitatively, as shown in Fig. 3f in the main text.

Supplemental references:

[1] J. Wu, H. Yuan, M. Meng, C. Chen, Y. Sun, Z. Chen, W. Dang, C. Tan, Y. Liu, J. Yin, Y. Zhou, S. Huang, H. Q. Xu, Y. Cui, H. Y. Hwang, Z. Liu, Y. Chen, B. Yan, and H. Peng, Nature Nanotechnology **12**, 530 (2017).